\documentclass[aps,prb,twocolumn,superscriptaddress,showpacs]{revtex4}
\usepackage{epsfig}
\usepackage{amsmath}
\usepackage[dvipsnames,usenames]{color}

\newcommand\mycite[1]{\raisebox{-0.5em}{\Large\cite{#1}}}
\begin{document}

\title{Spatial interferences in the electron transport of heavy-fermion materials}

\author{Shu-feng Zhang}
\affiliation{Beijing National Laboratory for Condensed Matter Physics and
Institute of Physics, Chinese Academy of Sciences, Beijing 100190, China}
\author{Yu Liu}
\affiliation{LCP, Institute of Applied Physics and Computational Mathematics, Beijing 100088, China}
\affiliation{Software Center for High Performance Numerical Simulation, China Academy of Engineering Physics, Beijing 100088, China}
\author{Hai-Feng Song}
\affiliation{LCP, Institute of Applied Physics and Computational Mathematics, Beijing 100088, China}
\affiliation{Software Center for High Performance Numerical Simulation, China Academy of Engineering Physics, Beijing 100088, China}
\author{Yi-feng Yang}
\email[]{yifeng@iphy.ac.cn}
\affiliation{Beijing National Laboratory for Condensed Matter Physics and
Institute of Physics, Chinese Academy of Sciences, Beijing 100190, China}
\affiliation{Collaborative Innovation Center of Quantum Matter, Beijing 100190, China}
\affiliation{School of Physical Sciences, University of Chinese Academy of Sciences, Beijing 100190, China}

\begin{abstract}
The scanning tunneling microscopy/spectroscopy and the point contact spectroscopy represent one of the major progresses in recent heavy fermion research. Both have revealed important information on the composite nature of the emergent heavy electron quasiparticles. However, a detailed and thorough microscopic understanding of the similarities and differences in the underlying physical processes of these techniques is still lacking. Here we study the electron transport in the normal state of the periodic Anderson lattice by using the Keldysh nonequilibrium Green's function technique. In addition to the well-known Fano interference between the conduction and $f$-electron channels, our results further reveal the effect of spatial interference between different spatial paths at the interface on the differential conductance and their interesting interplay with the band features such as the hybridization gap and the Van Hove singularity. We find that the spatial interference leads to a weighted average in the momentum space for the electron transport and could cause suppression of the electronic band features under certain circumstances. In particular, it reduces the capability of probing the $f$-electron spectral weight near the edges of the hybridization gap for large interface depending on the Fermi surface of the lead. Our results indicates an intrinsic inefficiency of the point contact spectroscopy in probing the $f$-electrons.
\end{abstract}

\maketitle

\section{Introduction}

Heavy electron materials with their rich emergent phenomena provide an important laboratory playground for studying the underlying organizing principles of strongly correlated electrons \cite{Stewart1984, Degiorgi1999, HandbookMagnetism}. One of the important recent progresses is the prediction of the Fano interference in the scanning tunneling microscopy/spectroscopy (STM/STS) and the point contact spectroscopy (PCS), first in Ref. \mycite{YangPRB2009} and then elaborated in a number of papers\cite{Fogelstrom01,Fogelstrom02,Fogelstrom03,Maltseva2009PRL,Wolfle2010PRL, FigginsPRL,FigginsPRL2011,Dubi2011PRL, JXZhu2012PRL,Yuan2012PRB,Toldin2013PRB,Dzero2014JPSJ}. The typical Fano lineshape in the conductance spectrum has now been observed in a number of heavy-fermion materials including CeCoIn$_5$ \cite{CeCoIn5Park2008PRL,CeCoIn5Aynajian,Aynajian2014JPSJ,JPCS(Park2009),CeCoIn5Davis,CeCoIn5STM(Zhou)}, URu$_2$Si$_2$ \cite{URu2Si2Schmidt2010,URu2Si2Aynajian2010PNAS,URu2Si2ParkPRL}, YbRh$_2$Si$_2$ \cite{YbRhSi2011Ernst}, YbAl$_3$ \cite{YbAl3Park}, and the recently proposed topological Kondo insulator SmB$_6$ \cite{XHZhangPRX,SmB6Hoffman,SmB6WangPRL,SmB6PNAS2014}. These works confirm the microscopic origin of the emergent heavy electron quasiparticles as the composition of conduction electrons and localized $f$-spins. However, despite of these important progresses, detailed analysis of experimental data has proved difficult, because of the lack of a thorough microscopic understanding of the underlying physical processes behind these techniques, not to mention the strongly correlated nature of the heavy-electron physics and the complex band properties. In particular, it is known experimentally that STS and PCS yield very different conductance spectra for the same individual material and cause difficulty and ambiguity in the interpretation of the experimental data and the underlying physics. For example, while the typical feature of the hybridization gap has been generally observed in STS, it has been missing in many PCS measurements such as CeCoIn$_5$ in the normal metallic state \cite{CeCoIn5Park2008PRL}. In URu$_2$Si$_2$, it remains unclear if the observation of the hybridization gap in the PCS is necessarily associated with the hidden order state \cite{Dubi2011PRL,URu2Si2Schmidt2010,URu2Si2Aynajian2010PNAS,URu2Si2ParkPRL,URu2Si2Lu2012}. In contrast to STM/STS, it is also highly debated whether PCS can provide useful information related to the electronic band structures \cite{Harrison1961,LeePCS2014,LeePCS2015}.

\begin{figure}[b]
  \begin{center}
    \includegraphics[width=1\columnwidth]{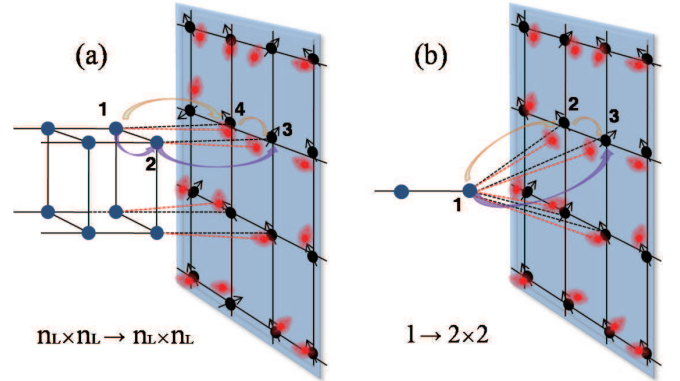}
  \caption{(Color online) An illustration of the simplified experimental setup for different configurations: (a) $n_L\times n_L\to n_L\times n_L$ with $n_L=2$ here and (b) $1\to2\times2$. The arrows indicate different pathways for the electron to inject into the lattice that cause spatial interference and modulation to the differential conductance.}
  \label{figsetup}
  \end{center}
\end{figure}

In this work, we study the physical processes underlying both techniques by using a much simplified toy model under the large-$N$ mean-field approximation. We discretize the lead and the interface between the periodic Anderson lattice and the metallic lead \cite{Berthod2011}. The overall setup is illustrated in Fig. \ref{figsetup}. The model has neglected many of the properties of a real heavy fermion system and may not represent the realistic situation of a scanning tunneling or point contact experiment, but it allows us to capture some of the essential features of the underlying physics such as the typical hybridization gap of the heavy fermion system and the interference effect of the electron transport through the interface. Moreover, by gradually increasing the size of the interface, we can see how the conductance spectra evolve from the case of a local metallic tip to a point contact of finite size, which may reveal some of the similarities and differences between STS and PCS in a very crude approximation.

As a matter of fact, our model reveals not only the band features such as the hybridization gap and the well-known Fano interference, but also additional spatial interference effect for electrons entering the lattice through different paths at the interface. We find that the conductance spectra are highly modulated by the pattern of this spatial interference. For example, the edges of the hybridization gap that are clearly seen in the $1\to 1$  configuration could be smoothed out for larger contact, causing possible confusion between the hybridization gap and the typical destructive interference in the Fano effect, both of which are seen as a broad minimum in the conductance spectra. The Van Hove singularity originated from the two-dimensional (2D) lattice model could also be completely suppressed in the $1\to 2\times 2$ configuration. The difference between $1\to 1$ and $1\to 2\times 2$ configurations may be connected to the STM measurement on URu$_2$Si$_2$, where one may find different conductance spectra for the tip located right on top of an $U$-site or in the middle of a cell \cite{URu2Si2Aynajian2010PNAS}. In addition, we find that the spatial interference may also lead to asymmetry in the conductance spectrum with respect to the bias voltage. This may be understood as a weighted average in the momentum space and prevent a faithful probe of the local electronic density of states in PCS, in contrast to the case of STS. For heavy-fermion materials, this leads to a constraint on the momentum space and suppresses the probe of the $f$-electron spectral weight in PCS, which we believe is one of the major differences between the two techniques. On the other hand, it also indicates that PCS has the potential to detect certain changes in the Fermi surface that cannot be averaged out, as studied in Ref. \mycite{LeePCS2014}.

This paper is organized as follows. Section II describes in detail the experimental setup and the model Hamiltonian, where we derive the current formula by using the Keldysh nonequilibrium Green's function technique and solve the model with numerical simulations under the large-$N$ mean field approximation. Section III presents our main numerical results and discuss the implication of these results. We will focus on the effect of the Fano interference and the spatial interference and their interplay with the band features including the hybridization gap and the Van Hove singularity. In Sec. IV, we discuss briefly the validity and limitation of our approach.

\section{The Model}

We start by considering a much simplified model for the experimental setup as illustrated in Fig. \ref{figsetup}. The whole setup includes a periodic lattice of the heavy fermion material under consideration and a metallic lead that is modelled by a discretized quasi-one-dimensional chain. The lead forms a junction with the heavy fermion lattice at one end and extends to infinity at the other end. In this work, the junction is assumed to have a square cross section of the size of $n_L\times n_L$ on the lead side which are then coupled to a chosen number of the lattice sites. Figure \ref{figsetup} illustrates two  different configurations with a junction of $2\times 2 \to 2\times 2$ and $1\to 2\times 2$, respectively. Thus we may tune the size of the cross section and the couplings to control the number of transport channels to resemble the change from a local STM tip to the point contact and study the gradual change of the electron conductance in the combined system. For simplicity, we have neglected all possible complications such as diffusion or energy relaxation, surface reconstruction, or scattering due to boson excitations at the interface. This approximation allows us to model the system in a simple way using the standard technique of quantum transport. In this sense, our model may be considered as a "quantum" point contact. We note that this much simplified model may not be completely appropriate in describing the realistic point contact experiment. However, as we will show below, it could indeed capture some of the basic physics of the underlying processes and provide us with interesting insight concerning both techniques for a better understanding of the similarities and differences between STS and PCS.

\subsection{The Hamiltonian}
We model the system using the following microscopic Hamiltonian that contains three parts,
\begin{equation}
\hat H=\hat H_L+\hat H_S + \hat H_t,
\end{equation}
where $\hat H_L$, $\hat H_S$ and $\hat H_t$  describe the metallic lead ($L$), the heavy fermion system ($S$), and their coupling at the interface ($t$), respectively. The lead is then approximated as a semi-infinite quasi-one-dimensional chain with a square cross section of $n_L\times n_L$ lattice sites described by
\begin{eqnarray}
\hat H_{L}=- t_L\sum_{\langle ij\rangle\sigma} \hat a_{i\sigma}^\dag \hat a_{j\sigma},
\label{hamiltonian1}
\end{eqnarray}
in which the operator $\hat a_{i\sigma}^\dag$ ($\hat a_{i\sigma}$) creates (annihilates) an electron at site $i$ with spin $\sigma$ inside the lead. We make further assumptions that the electrons in the lead are noninteracting. We will see that the exact model of the lead will not affect the qualitative conclusion on the properties of the electron transport in the combined system. The coupling Hamiltonian between the lead (left) and the heavy fermion lattice (right) is written as,
\begin{eqnarray}
   \hat H_{t}=\sum_{\langle lis\rangle\sigma}(t_{s} \hat a_{l\sigma}^\dag \hat c_{is\sigma}+\text{H.c.}),
\end{eqnarray}
which describes the electron hopping into the lattice. $\hat c_{is\sigma}$ is the annihilation operator of the lattice electron at site $i$ with spin $\sigma$ and orbit $s$ (corresponding to the conduction or $f$-band). $\hat H_S$ is introduced later. Here the Hamiltonian is written in real space and the hopping is only allowed between adjacent lattice points within the contact region of the junction. The detailed form of the hopping matrix may be very complicated depending on the specific geometry of the contact. For example, in the case of STM, the tip is very sharp so that only one local site at the tip is considered and the hopping of the electrons from the tip to the nearest lattice site may be modelled as a $1\to1$ configuration. On the other hand, if the tip is located near the center of a unit cell of a square lattice, then the electrons on the tip may have equal probability to hop to its four neighboring lattice sites and this should be modelled as a $1\to 2\times 2$ configuration. In the case of PCS, there may exist hundreds of small contacts on the interface, and each small contact may cover a large number of lattice sites \cite{PCSreview2005,rewPCS(Daghero),rewPCS(Park)}, which may instead be modelled as a $n_L\times n_L\to n_L\times n_L$ configuration with a large $n_L$. For simplicity, we neglect the details of the geometry of the contact region and assume constant hopping parameters for each orbit $s$. Figure \ref{figsetup} illustrates two possible configurations corresponding to the $n_L\times n_L\to n_L\times n_L$ case and the $1\to n_L\times n_L$ case with $n_L=2$.

\subsection{The Current Formula}

Now we derive the current formula for the above setup using the Keldysh nonequilibrium Green's function formalism. The electron current is given by \cite{Wingreen,Wingreen1994},
\begin{eqnarray}
I&=&-2e\langle\partial_t \hat N_L\rangle=2\frac{ie}{\hbar}\langle[\hat N_L,\hat H]\rangle\nonumber\\
&=&2\frac{ie}{\hbar}{\rm Tr}\langle \hat a_L^\dag \check H_{LS}\hat c_S-\hat c_S^\dag \check H_{SL}\hat a_L\rangle,
\label{Eqc4}
\end{eqnarray}
where $2\hat N_L=2\sum_{i\in L} \hat a_i^\dag \hat a_i$ is the number operator of the electrons in the lead. The prefactor 2 accounts for the spin degeneracy and the spin indices are dropped for simplicity. $\check H_{LS}$ is the hopping matrix from the lead to the lattice system determined by the hopping parameters $t_{s}$ defined in $\hat H_t$, where the symbol ``$\check{\phantom{a}}$'' indicates a matrix. The operators $\hat c_S=(...,\hat c_{is},...)^T$ and $\hat a_L=(...,\hat a_l,...)^T$ include all the lattice/lead sites that are coupled through the hopping matrix.

To proceed, we define the lesser Green's function, $\check G_{SL}^<(t)=i\langle \hat a_L^\dag(0) \hat c_S(t)\rangle$. The current formula is then rewritten as
\begin{eqnarray}
I&=&\frac{2 e}{\hbar}{\rm Tr}(\check H_{LS}\check G_{SL}^<(t=0) -\check H_{SL}\check G_{LS}^<(t=0) )  \nonumber \\
&=&\frac{2 e}{h}{\rm Tr}\int d\omega[\check H_{LS}\check G_{SL}^<(\omega) -\check H_{SL}\check G_{LS}^<(\omega) ].
\end{eqnarray}
Using the Dyson equation and the Langreth theorem \cite{bookHaug}, we have
\begin{eqnarray}
\check G_{SL}^<=\check G_{SS}^r \check H_{SL}\check g_{LL}^<+\check G_{SS}^< \check H_{SL}\check g_{LL}^a\nonumber,\\
\check G_{LS}^<=\check g_{LL}^r \check H_{LS}\check G_{SS}^<+\check g_{LL}^< \check H_{LS}\check G_{SS}^a,
\end{eqnarray}
where $\check g_{LL}^{r,a,<}$ are the retarded, advanced, or lesser Green's functions of the lead alone without the coupling to the lattice, whereas $\check G_{SS}^{r,a,<}$ are the corresponding full Green's functions of the lattice that includes the effect of the coupling to the lead. The self-energy for the lattice electrons due to this coupling is given by,
\begin{eqnarray}
\check \Sigma_L^{r,a,<}\equiv \check H_{SL}\check g_{LL}^{r,a,<}\check H_{LS}.
\label{EqSigmaL}
\end{eqnarray}
The current formula then becomes
\begin{eqnarray}
I&=&\frac{2e}{h}{\rm Tr}\int d\omega [\check \Sigma_L^<(\check G_{SS}^r-\check G_{SS}^a)-
\check G_{SS}^<(\check \Sigma_L^r-\check \Sigma_L^a )].\nonumber\\
\label{currentformula4}
\end{eqnarray}
On using the fluctuation-dissipation theorem, we have
\begin{equation}
\check \Sigma_L^<=-f_L(\check \Sigma_L^r-\check \Sigma_L^a )=if_L\check \Gamma_L,
\end{equation}
where $\check \Gamma_L=i(\check \Sigma_L^r-\check \Sigma_L^a)$ represents the broadening of the spectral function of the lattice electrons due to the coupling to the lead. Hence
\begin{eqnarray}
I=\frac{2e}{h}{\rm Tr}\int d\omega [if_L\check \Gamma_L(\check G_{SS}^r-\check G_{SS}^a)+i\check \Gamma_L
\check G_{SS}^<],
\label{currentformula3}
\end{eqnarray}
where $f_{L}=f(\omega-eV)$ is the Fermi distribution function of the electrons in the lead shifted by the  bias voltage $V$.

To calculate the lattice Green's functions, we use the Dyson equation,
\begin{eqnarray}
 \check G_{SS}^{r/a-1}=&\check g_{SS}^{r/a-1}-\check \Sigma_{L}^{r/a},
  \label{EqGg}
\end{eqnarray}
where $\check g_{SS}^{r/a}$ is the Green's function of the lattice alone without the coupling to the lead and the effect of the lead is fully incorporated in the self-energy matrix. The above formula is exact for a noninteracting system. In general, the coupling may have more subtle effects on the lattice Green's function, but these are all higher-order contributions and can be safely neglected in our case for a qualitative study of the conductance spectrum of the heavy fermion state.

The lesser Green's function in Eq. (\ref{currentformula3}), however, has a more complicated form (see Appendix \ref{appendixCformula}),
\begin{eqnarray}
\check G_{SS}^<=-f_S(\check G_{SS}^r-\check G_{SS}^a)
-i(f_S-f_L)\check G_{SS}^r\check \Gamma_{L}\check G_{SS}^a,
\label{EqGle}
\end{eqnarray}
where $f_S=f(\omega)$ is the Fermi distribution function in the lattice. In addition to the common (first) term that may be derived from the fluctuation-dissipation theorem for the lattice system, we find in the above equation a new (second) term that is associated with the difference in the Fermi distribution functions of the lead and the lattice, $f_S-f_L$. It is therefore a genuine nonequilibrium effect and originates from the redistribution of the electron energy spectra at the interface due to the difference in the chemical potentials, which indicates that the combined system is not in the equilibrium state and  the usual fluctuation-dissipation theorem does not apply.

Using the above results, we can eventually rewrite the current formula as
\begin{eqnarray}
I=\frac{2e}{h} \int d\omega(f_{L}-f_{S}){\rm Tr }[i \check \Gamma_{L}(\check G_{SS}^r(\omega)-\check G_{SS}^a(\omega)) \nonumber\\
-\check \Gamma_{L} \check G_{SS}^r(\omega) \check \Gamma_{L} \check G_{SS}^a(\omega)],
\label{currentformula1}
\end{eqnarray}
where all the Green's functions are written in a matrix form in real space. At zero temperature, the differential conductance is given by
\begin{eqnarray}
G(\omega)=\frac{2e^2}{h}{\rm Tr }[i\check \Gamma_{L}(\check G_{SS}^r(\omega)-\check G_{SS}^a(\omega))\nonumber\\
- \check \Gamma_{L} \check G_{SS}^r(\omega) \check \Gamma_{L} \check G_{SS}^a(\omega) ].
\label{Eqconductance01}
\end{eqnarray}
The first term in the above formula is commonly seen in the literature \cite{YangPRB2009,Maltseva2009PRL}, whereas the second term is usually not present and in our derivation comes from the second term in the lesser Green's function of the lattice in Eq. (\ref{EqGle}). It is of higher order in $\check \Gamma_L$ and usually not important in the weak coupling regime with $t_s$ or $\check \Gamma_L$ much smaller than the typical energy scale of the lattice electrons. However, as we will discuss later, in the strong coupling regime, it cannot be neglected and may in fact play a crucial role and cause some unexpected results in the conductance spectrum. As discussed above, this second term originates from the local nonequilibrium effect that modifies the electron distribution at the interface of the lead and the lattice. This change in the electron distribution provides a negative feedback and reduces the probability for the electron transmission. As shown in Appendix \ref{appendixScattering}, our formula is consistent with the well-known Landauer formula \cite{book(Datta)}.

\subsection{The Anderson lattice}
Heavy fermion systems are typically described by the periodic Anderson model, which is, however, a strongly correlated model and cannot be solved exactly. For simplicity, we consider the large-$N$ limit \cite{BickersRMP1987} ($N=2$), where, neglecting the finite self-energy, the low-energy quasiparticles can be approximately described by an effective mean-field Hamiltonian,
\begin{equation}
\hat H_S=\sum_{\mathbf k\sigma} \hat\varphi_{\mathbf k\sigma}^\dag
\check H_S(\mathbf k)\hat\varphi_{\mathbf k\sigma},
\end{equation}
where $\hat\varphi_{\mathbf k\sigma}=(\hat c_{{\mathbf k}\sigma}, \hat f_{{\mathbf k}\sigma})^T$ and $\hat c_{{\mathbf k}\sigma}$ and $\hat f_{{\mathbf k}\sigma}$ are the annihilation operator of the conduction electron and the $f$-electron of momentum ${\mathbf k}$ and spin $\sigma$, respectively. We have
\begin{eqnarray}
\check H_s({{\mathbf k}})=\left(\begin{array}{cc}
  \epsilon_{c{\mathbf k}}&\mathcal{V}\\\mathcal{V}&\epsilon_f
\end{array}\right),
\end{eqnarray}
in which the dispersion of conduction band is $\epsilon_{c{\mathbf k}}=-\mu-2t({\rm cos}(k_x)+{\rm cos}(k_y))$ for a square lattice and $\mathcal{V}$ is the hybridization between the conduction band and the renormalized flat $f$-electron band located at the energy $\epsilon_f$. This yields two hybridization bands whose dispersion can be immediately written down as
\begin{eqnarray}
  E_{\pm}({\mathbf k})=\frac{1}{2}\left[(\epsilon_{c{\mathbf k}}+\epsilon_{f})
  \pm\sqrt{(\epsilon_{c{\mathbf k}}-\epsilon_{f})^2+4\mathcal{V}^2}\right],
\label{Eqbandn}
\end{eqnarray}
with an indirect hybridization gap given by $\Delta_h\approx\frac{2\mathcal{V}^2}{D}$, where $D=4t>>\mathcal{V}$ corresponds to the half width of conduction band $\epsilon_{c{\mathbf k}}$ and is the largest energy scale in this study. For the 2D square lattice, we have also the Van Hove singularity at ${\mathbf k}=(0,\pm\pi)$ and $(\pm\pi,0)$ in the conduction band, which leads to divergence in the density of states of the hybridization bands at the energies $E_{\pm}^{VH}=\frac{1}{2}[(-\mu+\epsilon_{f}) \pm\sqrt{(\mu+\epsilon_{f})^{2}+4\mathcal{V}^{2}}]$, respectively. These band features, the hybridization gap and the Van Hove singularity, determine the primary shape of the conductance spectrum.

Within the mean-field approximation, we can write analytically the Green's functions of the lattice electrons in the momentum space in the matrix form,
\begin{equation}
\check g_{SS}^r({\mathbf k},\omega)=\langle\langle \hat\varphi_{\mathbf k\sigma}| \hat\varphi_{\mathbf k\sigma}^\dag \rangle\rangle^r=\left[\omega\check I- \check H_{S}({\mathbf k})+i0^+\right]^{-1},
\end{equation}
where $\check I$ is the unit matrix. For later use, we give the analytical form of all components of the above Green's functions in Appendix \ref{appendixGreen}. They can then be Fourier transformed into the real space and combined with Eqs. (\ref{EqSigmaL}) and (\ref{EqGg}) to obtain the full Green's functions of the coupled system, $\check G^{r,a}_{SS}(\omega)$, in order to calculate the electric current using Eq. (\ref{currentformula1}). We note that the corresponding hopping parameters $t_s$ may also be renormalized by the electronic correlations, which will be taken as a free input parameter for simplicity in this work. We use $t_c$ and $t_f$ to denote the hopping parameters to the conduction electrons and the $f$-electrons, respectively, and further assume that both are constant.

\section{Numerical results}

We now discuss in detail the resulting conductance spectra calculated for different configurations and input parameters using the above model. We consider in particular the following configurations: $1\to 1$, $1\to 2\times 2$, and $n_L\times n_L\to n_L\times n_L$ with $n_L=2, 4, 8$ and $16$. Because of the coupling to the lead, the planar translational symmetry of the lattice is broken, so that the self-energy $\check \Sigma_L^r$ and the full Green's functions, $\check G^{r,a}_{SS}(\omega)$, can only be calculated numerically in real space \cite{DHLsgf,Rubio1984,Rubio1985}. In the following, we focus first on the variation of the conductance spectra as a function of the ratio $q=t_f/t_c$ in the weak coupling ($t_{f,c}<<t$) case, which is the typical situation for STS. The strong coupling ($t_{f,c}\sim t$) case is discussed briefly at the end of this section. In the case of PCS, the coupling strength  $t_{f,c}$ could be larger because of the direct contact. However, their values actually depend on the overlap with the electron orbitals in the leads and those on the cleaved surface of the lattice. For example, $t_c$ could be small if the cleaved surface is not in the conduction plane that is responsible for the hybridization. On the other hand, as we will show later, the magnitude $t_{f,c}$ only becomes important for the STM case ($n_L=1$). Its effect is suppressed for PCS with large $n_L$. Therefore, for direct comparisons, we study both parameter ranges to see clear the evolution as a function of the size of the contact. Our results reveal important interference effects for electron transport through different channels (the conduction and $f$-electron bands) and different paths (different sites at the interface). These lead to the interplay between the so-called Fano interference and the spatial interference. We further show that the spatial interference is equivalent to a weighted average in the momentum space of the lattice electrons. Both effects may interplay with the band properties such as the Van Hove singularity and the hybridization gap to determine the overall features of the conductance spectra.

\subsection{The Fano interference}

\begin{figure}[t]
  \begin{center}
   \includegraphics[width=\columnwidth]{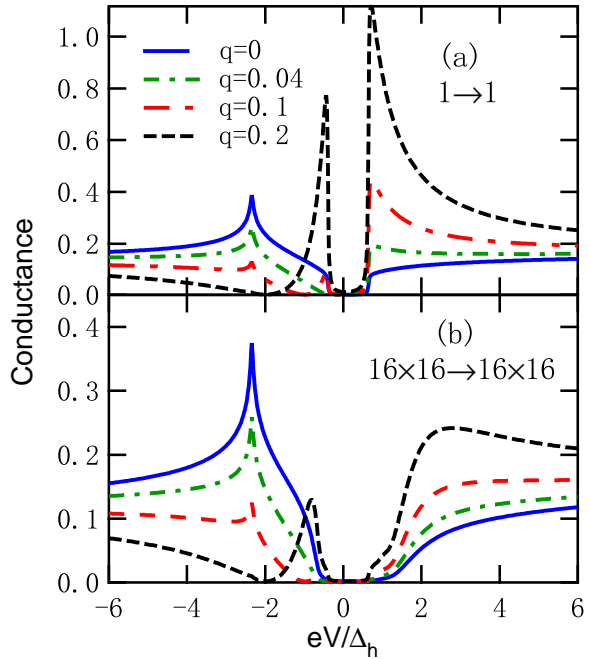}
  \caption{(Color online) Conductance spectra for (a) $1\to1$ and (b) $16\times16\to16\times16$ configurations with $q=0$, 0.04, 0.1 and 0.2. The parameters are $t=25$, $\mu=-20$, $\epsilon_f=0.01$, $\mathcal{V}=5$, $t_{c}=0.001t$, $t_L=t$, $\Gamma=0.01$. The bias voltage is normalized by the width of the hybridization gap, $\Delta_h=2\mathcal{V}^2/D$ and the conductance is in units of $\frac{2e^2}{h}n_L^2\frac{4\pi t_{c}^2}{t_L t}$.  }
  \label{figfano01}
  \end{center}
\end{figure}

We first calculate the differential conductance for the $1\to1$ configuration. It is now well-known that electrons tunneling through both the conduction and $f$-electron channels exhibit the so-called Fano interference \cite{reviewFano,YangPRB2009}. Figure \ref{figfano01}(a) shows the variation of the calculated spectra with the Fano parameter, $q=t_f/t_c$, whose value determines the exact location of the conductance minimum due to the Fano destructive interference. For $q=0$, the tip couples locally only to the conduction band so that the spectrum follows exactly the density of states of the conduction electrons and displays clearly the Van Hove peak and the hybridization gap. However, with increasing $q$, a complicated change is seen in the overall features of the spectra. Whereas the right edge is always enhanced and gets sharper with increasing $q$ due to the increasing probability of tunneling into the $f$-electron channel whose spectral weight peaks at the edge of the hybridization gap, the left edge of the hybridization gap is first suppressed for $q=0.04$, and then reappears for larger values of $q$. In the mean time, the Van Hove peak is gradually suppressed and completely disappears at $q=0.2$. This nonmonotonic variation of the left edge and the suppression of the Van Hove peak originate from the shift of the Fano minimum in the conductance and manifest an interesting interplay of the Fano destructive interference with the hybridization gap and the Van Hove singularity. We note that in all the figures, we set arbitrarily the zero energy point to be inside the hybridization gap.

To see this more clearly, we neglect the higher order term in Eq. (\ref{Eqconductance01}) and rewrite the conductance formula as,
\begin{eqnarray}
G(eV)=\frac{2e^2}{h}(2\pi)^2 t_{c}^2 {\rm Tr}\left[\check \rho_{L}(0)(\check \rho_c\!+\!2q\check \rho_{cf}\!+\! q^2\check\rho_f)_{eV}\right],
\label{Eqconductance03}
\end{eqnarray}
where $\check \rho_L(0)=-{\rm Im}\check g_{LL}^r(\omega=0)/\pi$ is the imaginary part of the lead's Green's function at the Fermi energy of the lead at the interface in real space, and $\check \rho_{c}=-{\rm Im}\check g_{cc}^r(eV)/\pi$, $\check \rho_{cf}=-{\rm Im}\check g_{cf}^r(eV)/\pi$ and $\check \rho_{f}=-{\rm Im}\check g_{ff}^r(eV)/\pi$ are those of the lattice Green's functions at the corresponding energy with a finite bias voltage of $V$. The trace is over all the lattice sites at the interface. For $1\to 1$ configuration, there is only one site at the interface so the above quantities are all real numbers and represent the density of states of the corresponding components. For $q=0$, we see that the differential conductance is simply in proportion to the density of states, $\rho_c(eV)$, of the lattice conduction electrons. For finite $q$, there are also contributions from the direct tunneling into the $f$-electron band, $\rho_f$, and the mixed term, $\rho_{cf}$, which reflects the interference effect because of the hybridization. As shown in Appendix \ref{appendixGreen}, in the mean-field approximation, the above formula can be further simplified as
\begin{eqnarray}
G(eV)=\frac{2e^2}{h}(2\pi)^2 t_{c}^2{\rm Tr}\left[\check\rho_{L}(0) \check\rho_c(eV)\right]\mathcal{F}(eV),
\label{Eqconductance04}
\end{eqnarray}
in which the term $\mathcal{F}(eV)=\left(1+q\frac{\mathcal{V}}{eV -\epsilon _{f}}\right)^{2}$ has the typical Fano lineshape, while the term ${\rm Tr}\left[\check \rho_{L}(0) \check \rho_c(eV)\right]$ contains the band features, namely the hybridization gap and the Van Hove singularity. In this case, we see that the Fano interference shows up as a simple modulation to the overall spectra. The location of the conductance minimum is then located at $eV=\epsilon_f-q\mathcal{V}$, which moves gradually to lower energies with increasing $q$ and suppresses in sequence the left edge of the hybridization gap at $q=0.04$ and the Van Hove peak at $q=0.2$.

This Fano feature and its interplay with the hybridization gap and the Van Hove singularity are also seen in the $16\times16\to16\times16$ configuration as shown in Fig. \ref{figfano01}(b). The above formula also applies in this general situation, but $\check \rho_L$ and $\check \rho_c$ are no longer given by the local density of states. They involve off-diagonal elements that give rise to spatial interference for electrons entering the lattice through different paths or sites at the interface. As a result, the edges of the hybridization gap are further smoothed out, as clearly seen in our numerical results. This interesting interplay between the Fano interference and the band features has not been discussed explicitly in previous literatures and may cause misinterpretation of the experimental data.

\subsection{The spatial interference}

\begin{figure}[t]
  \begin{center}
  \includegraphics[width=\columnwidth]{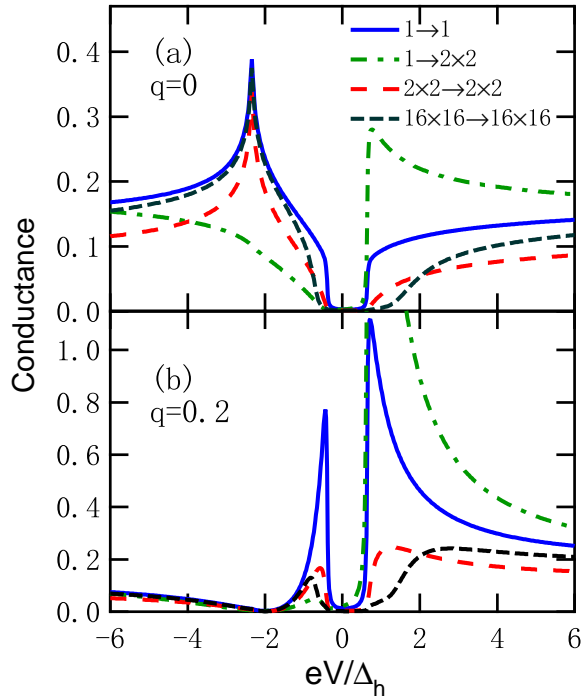}
  \caption{(Color online)  Conductance spectra for (a) $q=0$ and (b) $q=0.2$ for different experimental configurations. The parameters are $t=25$, $\mu=-20$, $\epsilon_f=0.01$, $\mathcal{V}=5$, $t_{c}=0.001t$, $t_L=t$, $\Gamma=0.01$. The conductance is in units of $\frac{2e^2}{h}n_R^2\frac{4\pi t_{c}^2}{t_L t}$ with $n_R=n_L$ for $n_L\times n_L\to n_L\times n_L$ and $n_R=2$ for $1\to2\times2$ configuration.  }
  \label{figpath01}
  \end{center}
\end{figure}

We now discuss the spatial interference effect in more details by studying different hopping configurations. Figures \ref{figpath01}(a) and \ref{figpath01}(b) show our results for four configurations ($1\to 1$, $1\to 2\times2$, $2\times 2\to 2\times 2$, and $16\times 16 \to 16\times 16$) with $q=0$ and 0.2, respectively. For $q=0$, where electrons only enter the conduction channel, the Van Hove peak is qualitatively unchanged for all $n_L\times n_L\to n_L\times n_L$ configurations with increasing $n_L$ from 1 to 16, whereas the two edges of the hybridization gap are gradually smeared out so that the gap becomes a broad valley, which may be easily confused with the Fano destruction minimum typically seen in the point contact spectra of heavy-fermion materials \cite{CeCoIn5Park2008PRL,URu2Si2ParkPRL,XHZhangPRX}. Interestingly, however, the spectrum exhibits completely different pattern in the $1\to2\times2$ configuration, which may be realized in the STM measurement. Whereas the Van Hove singularity is completely suppressed, the upper edge of the hybridization gap is greatly enhanced, resembling that of a finite $q$ due to the tunneling into the $f$-electron band. The data show both destructive and constructive interference effects on different sides of the hybridization gap. Figure \ref{figpath01}(b) shows the results for $q=0.2$, where completely different conductance spectra are obtained. The Van Hove peak coincides with the Fano minimum and is therefore completely suppressed by the Fano effect for all configurations. On the other hand, the hybridization gap seems to be enhanced compared to those of $q=0$ due to the tunneling into the $f$-electron band, except for $1\to 2\times 2$ configuration, where the lower edge is somewhat suppressed due to the spatial interference. The combination of the Fano minimum and the hybridization gap leads to an isolated peak in between, which may be easily confused with a mid-gap state.

To better understand the effect of the spatial interference, we rewrite the formula of the differential conductance in Eq. (\ref{Eqconductance04}) by diagonalizing $\check \rho_L$ and $\check \rho_c$, which are defined on the interface. Since $\check \rho_c$ is obtained from the Green's function of the lattice system alone in the weak coupling limit, it can be diagonalized through the Fourier transformation to the momentum space,
\begin{eqnarray}
\check \rho_c({{\mathbf r}_i,{\mathbf r}_j})=\frac{1}{N_k}\sum_{{\mathbf k}}
e^{i{\mathbf k}\cdot ({\mathbf r}_i- {\mathbf r}_j)}\rho_{c{\mathbf k}},
\end{eqnarray}
where ${\mathbf r}_{i,j}$ denotes the two sites on the interface and $\rho_{c{\mathbf k}}$ is the density of states of the lattice conductance electrons at momentum ${\mathbf k}$ and the relative energy $eV$ (corresponding to the chemical potential of the lead). On the other hand, for small $n_L$, $\check \rho_L$ can be diagonalized straightforwardly using
\begin{eqnarray}
\check \rho_{L}({{\mathbf r}_j,{\mathbf r}_i})=\sum_{m=1}^{n_L^2}P^\dagger_{jm}\rho_{Lm}P_{mi},
\end{eqnarray}
where the subscript $m$ denotes the $m$-th channel of the lead after the diagonalization, $\rho_{Lm}$ is the density of states of the $m$-th channel on the interface at the Fermi energy of the lead, and $P_{mi}$ is the unitary transformation matrix. The conductance formula is then immediately rewritten as
\begin{eqnarray}
G(eV)=\frac{8\pi^2e^2}{h}t_c^2\sum_{{\mathbf k},m=1}^{n_L^2}
|\mathcal{A}_{m{\mathbf k}}|^2 {\rho}_{Lm}(0) \rho_{c{\mathbf k}}(eV)\mathcal{F}(eV),\nonumber\\
\label{G19}
\end{eqnarray}
in which
\begin{eqnarray}
  \mathcal{A}_{m{\mathbf{k}}}=\frac{1}{\sqrt{N_k}}\sum_{{\mathbf r}_i}P_{mi}e^{i{\mathbf{k}}\cdot {\mathbf r}_{i}},
\label{Eq0517_02}
\end{eqnarray}
assuming a constant coupling $t_c$ for all the sites on the interface. $\mathcal{A}_{m{\mathbf k}}$ contains all the interference effect due to the different paths for the electron transport through the interface and may have a more complicated form if $t_{c}$ or $t_f$ is site-dependent. For the $1\to1$ configuration, there is only one path so that $\mathcal{A}_{m{\mathbf k}}=1$, independent of the momentum. We see that the conductance probes a convolution of the density of states of the lattice electrons and the Fano interference. Whereas for $2\times2\to2\times2$ configuration, we find that two degenerate channels among all four channels in the lead cross the Fermi energy and yield $|\mathcal{A}_{m{\mathbf k}}|\propto |\sin(\frac{k_x\pm k_y}{2})|$, which becomes zero for $k_x\pm k_y=0$, $\pm2\pi$. In the particular lattice model considered here, these points include both the lower edge (corresponding to ${\mathbf k}=(\pm\pi,\pm\pi)$) and the upper edge (corresponding to ${\mathbf k}=(0,0)$) of the hybridization gap. The gap edges in the conductance spectra are hence smoothed out. In contrast, in the case of $1\to2\times2$ configuration, we obtain $|\mathcal{A}_{m{\mathbf k}}|\propto |{\rm cos}(k_x/2){\rm cos}(k_y/2)|$, which  vanishes for either $k_x=\pm \pi$ or $k_y=\pm\pi$, but is maximal at ${\mathbf k}=(0,0)$. This explains the suppression of the conductance at the lower edge of the gap and the enhancement at the upper edge of the gap. In the same way, we see that the Van Hove peak is suppressed because it appears at ${\mathbf k}=(0,\pm\pi)$ and $(\pm\pi,0)$.

The $n_L\rightarrow\infty$ limit provides an important example to examine the effect of the path interference, in some sense resembling the extreme situation of PCS with large contact \cite{LeePCS2014}. In this limit, the translational invariance in both planar directions requires $\check \rho_L({\mathbf r}_i,{\mathbf r}_j)=\rho_L({\mathbf r}_i-{\mathbf r}_j)$, so that it may also be diagonalized through a Fourier transformation to the momentum space of the lead electrons. We have then
\begin{eqnarray}
\check \rho_{L}({{\mathbf r}_j,{\mathbf r}_i})=\frac{1}{N_k}\sum_{\mathbf k}e^{i{\mathbf k}\cdot({\mathbf r}_j-{\mathbf r}_i)}\rho_{L{\mathbf k}},
\end{eqnarray}
namely, $P_{{\mathbf k}i}=N_k^{-1/2}e^{-i{\mathbf k}\cdot{\mathbf r}_i}$. This immediately leads to
\begin{eqnarray}
  \mathcal{A}_{\mathbf{k}^\prime,\mathbf{k}}=\delta_{\mathbf{k}^\prime,\mathbf{k}},
\label{Eq0517_03}
\end{eqnarray}
which indicates planar momentum conservation for electrons entering the heavy fermion lattice from the lead. This, together with the energy conservation, greatly limits the overall number of the allowed channels and causes a weighted average over the planar momentum. We believe this difference between the $1\to 1$ configuration and the $n_L\times n_L\to n_L\times n_L$ configuration (for large $n_L$) reflects one of the most fundamental distinctions between STS and PCS. The conductance formula in Eq. (\ref{G19}) then becomes
\begin{eqnarray}
G(eV)=\frac{8\pi^2e^2}{h}t_c^2\sum_{{\mathbf k}}{\rho}_{L{\mathbf k}}(0) \rho_{c{\mathbf k}}(eV)\mathcal{F}(eV).
\label{G23}
\end{eqnarray}
For an artificial 2D lead with noninteracting electrons, both $\rho_{L{\mathbf k}}$ and $\rho_{c{\mathbf k}}$ are $\delta$-functions. Hence the conductance is only nonzero if the energy conservation condition, $E_\pm({\mathbf k})-\epsilon_{L{\mathbf k}}=eV$, is met. The situation is slightly different in a realistic three-dimensional system. In this case, one may expect that the momentum, ${\mathbf k}_z$, along the perpendicular direction is not conserved. Hence the sum over ${\mathbf k}_z$ may modify the local density of states at the interface and change the electron spectral function from a simple $\delta$-form. The above strict constraint is then released. However, as we will see below, it still yields substantial modifications to the overall structure of the spectrum.

\begin{figure}[t]
  \begin{center}
   \includegraphics[width=1.05\columnwidth]{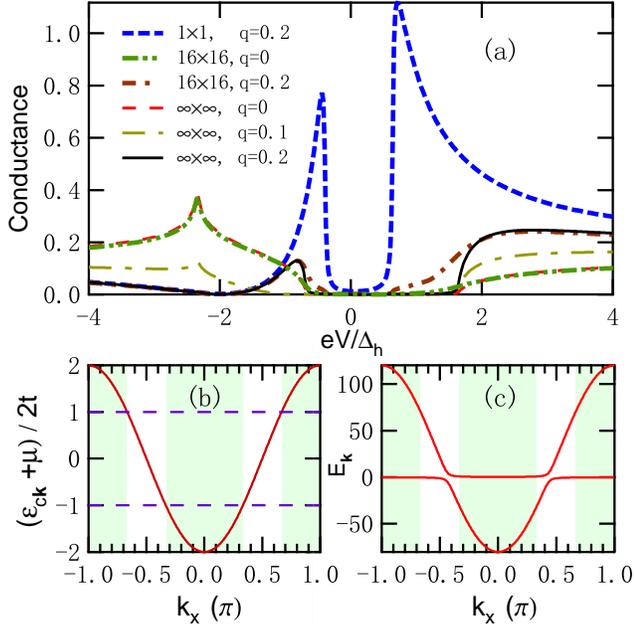}
  \caption{(Color online) (a) Comparison of the conductance spectra for $n_L=1$, 16 and $\infty$ configurations;
The parameters are $t=25$, $\mu=-20$, $\epsilon_f=0.01$, $\mathcal V=5$, $t_{c}=0.001t$, $t_L=t$, $\Gamma=0.01$. The conductance is in units of $\frac{2e^2}{h}n_L^2\frac{4\pi t_{c}^2}{t_L t}$. (b) and (c) Illustration of the momentum constraint along the $(1,1)$ direction for a noninteracting square lattice and the hybridization band of a heavy fermion lattice, respectively. The shaded regions do not contribute to the transport because of the constraint.}
  \label{Fig:N}
  \end{center}
\end{figure}

\begin{figure}[t]
  \begin{center}
   \includegraphics[width=\columnwidth]{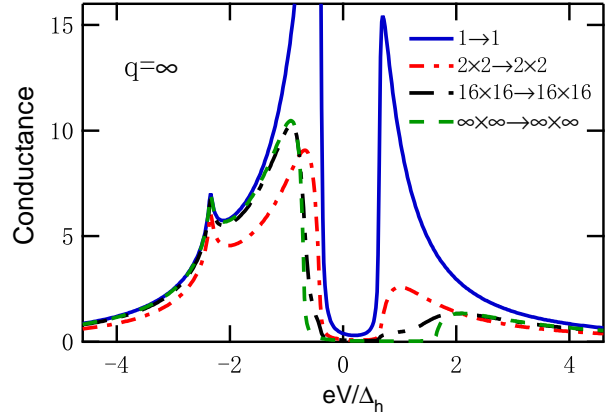}
  \caption{(Color online) Comparison of the conductance spectra at $q=\infty$ for several configurations with $n_L=1$, 2, 16 and $\infty$. The parameters are $t=25$, $\mu=-20$, $\epsilon_f=0.01$, $\mathcal V=5$, $t_{c}=0$, $t_L=t$, $\Gamma=0.01$, $t_f=0.001t$. The conductance is in units of $\frac{2e^2}{h}n_L^2\frac{4\pi t_{f}^2}{t_L t}$.  }
  \label{Fig:qinf}
  \end{center}
\end{figure}

As an example, Fig. \ref{Fig:N}(a) compares the calculated spectra for $n_L$=1, 16 and $\infty$ in our particular model which contains a semi-infinite lead coupled to a two-dimensional lattice. Compared to the results of $n_L=1$ and 16, the differential conductance for $n_L=\infty$ is surprisingly suppressed in a much wider range of the bias voltage compared to the hybridization gap, independent of the value of the Fano parameter $q$. This suppression reflects the particular form of the electron density of state of the semi-infinite lead at half filling,
\begin{equation}
\rho_{L\mathbf{k}}(\omega)  =\frac{1}{\pi t_L}\sqrt{1-(\omega-\epsilon
_{L\mathbf{k}})^{2}/4t_L^2},
\end{equation}
where $\epsilon_{L{\mathbf k}} =-2t_{L}(\cos {\mathbf k}_{x}+\cos {\mathbf k}_{y})$. For $\omega=0$, this yields a weighted constraint on the planar momentum, $|\cos{\mathbf k}_x+\cos{\mathbf k}_y|\le1$, for the electrons at the Fermi energy of the lead which enter the lattice system and contribute to the transport. Because of the momentum conservation in the planar direction, this would yield a similar constraint on the momentum of the lattice electrons and suppress the electron transport to the band edges, where $|(\epsilon_{c{\mathbf k}}+\mu)/2t| \propto |\cos{\mathbf k}_x+\cos{\mathbf k}_y| > 1$ for a square lattice of free conduction electrons, as illustrated in Fig. \ref{Fig:N}(b). Moreover, in the periodic Anderson model where the conduction band is coupled to a flat $f$-electron band, this constraint on the momentum space, when applied to the hybridization bands, would also suppress a large fraction of the $f$-electron spectral weight which typically dominates at the edges of the hybridization gap, as illustrated by the shaded regions in Fig. \ref{Fig:N}(c). This effectively results in the wide gap feature in the conductance spectra, as seen in the formula.
\begin{equation}
\lim_{n_L\rightarrow\infty}\frac{G(eV)}{n_L^2}=\frac{2e^{2}}{h}\frac{4\pi t_{c}^{2}}{t_{L}t}\rho_{c}
(eV)\sqrt{t^{2}-h(eV)^2}{\mathcal F}(eV),
\end{equation}
where $h(eV)=[eV-{\mathcal V^{2}}/(eV-\epsilon_{f})+\mu]/2$. This suppresses the differential conductance for $|h(eV)|\ge t$ and yields a gap of approximately $4\mathcal{V}^2/D \approx 2\Delta_h$, twice the size of the hybridization gap. We further note that the constraint has already taken effect at finite $n_L$. As may be seen in Fig. \ref{Fig:N}(a), the gap edges are already smoothed out at $n_L=16$. The suppression of the capability of probing the $f$-electron spectral weight is best seen in the case of $q=\infty$, namely $t_c=0$ so that electrons only enter the $f$-electron channel. The results are plotted in Fig.~\ref{Fig:qinf} and we see that the gap is gradually enhanced with increasing $n_L$, indicating that the $f$-electron spectral weight near the edges of the hybridization gap is no longer probed for large $n_L$.

\begin{figure}[t]
  \begin{center}
  \includegraphics[width=1.05\columnwidth]{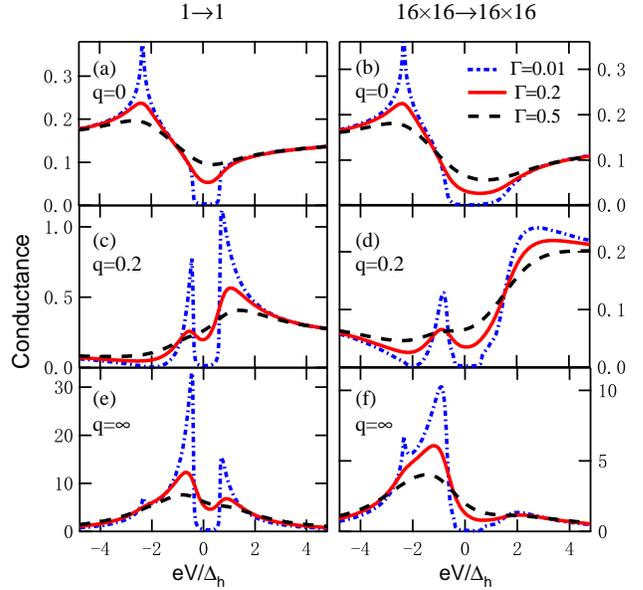}
  \caption{(Color online) Variation of the conductance spectra with $\Gamma$ for both $1\to1$ and $16\times16\to16\times16$ configurations with $q=0$, 0.2 and $\infty$. The parameters are $t=25$, $\mu=-20$,  $\epsilon_f=0.01$, $\mathcal{V}=5$, $t_L=t$. The conductance is in units of $\frac{2e^2}{h}n_L^2\frac{4\pi t_c^2}{t_Lt}$ for (a-d) with $t_c=0.001t$ and $\frac{2e^2}{h}n_L^2\frac{4\pi t_f^2}{t_Lt}$ for (e,f) with $t_f=0.001t$ and $t_c=0$.}
  \label{figs2_1}
  \end{center}
\end{figure}

The above constraint is not limited to our particular model. In general, the planar momentum for electrons at the Fermi energy is constrained by the projection of the lead's three-dimensional Fermi surface on the $(k_x, k_y)$ plane, or the range of the energy dispersion in the perpendicular direction, namely, $|\epsilon_{L{\mathbf k}}|\le W_z$, where, in our particular model, $W_z\approx 2t_L|\cos{\mathbf k}_z|=2t_L$. Hence, in contrast to STM or the $1\to 1$ configuration, which can probe the whole momentum space of the lattice electrons, the differential conductance in the case of large $n_L$, as may be the case of PCS, can only sample part of the momentum space depending on the Fermi surface of the lead. This, in combination with the energy conservation condition, may lead to "dark" regions in the Brillouin zone of the lattice electrons that do not satisfy the constraint and therefore cannot be probed. In the heavy fermion case, since the $f$-electron band is usually flat, the spectral weight of itinerant $f$-electrons spreads over the whole Brillouin zone and dominates around the edges of the hybridization gap. The effect of the "dark" region may then prohibit the electron transport into a large fraction of the $f$-band, as illustrated by the shaded regions in Fig. \ref{Fig:N}(c), and suppress the edges of the hybridization gap, leading to the broadening of the gap feature in the conductance spectra. This is a general conclusion from this study. We believe it is an intrinsic inefficiency of PCS in probing the $f$-electron spectral weight and may be crucial in understanding the experimental data.

\begin{figure}[t]
  \begin{center}
  \includegraphics[width=\columnwidth]{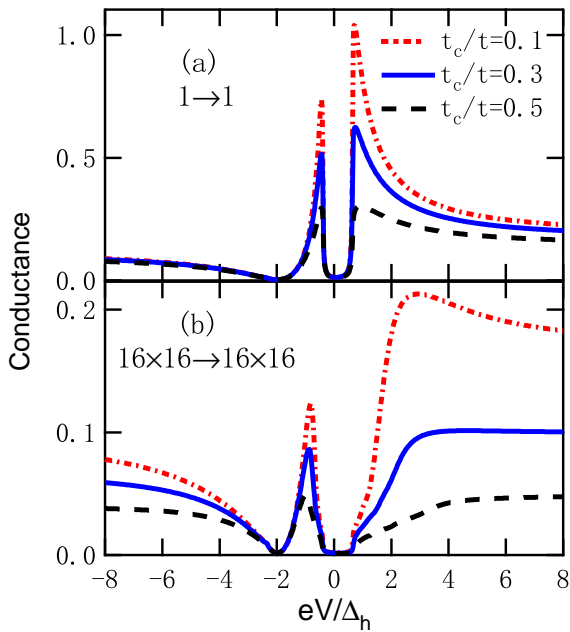}
  \caption{(Color online) The variation of the conductance spectra with $t_c$ for (a) $1\to1$ and (b)$16\times16\to16\times16$ configurations at $q=0.2$. The parameters are $t=25$, $\mu=-20$,  $\epsilon_f=0.01$, $\mathcal{V}=5$, $t_L=t$, $\Gamma=0.01$. The conductance is in units of $\frac{2e^2}{h}n_L^2\frac{4\pi t_c^2 }{t_Lt}$.  }
  \label{fig_ns}
  \end{center}
\end{figure}

In deriving the above results, we have, for simplicity, neglected the effect of electronic correlations on the lattice electrons. This may partly be accounted by introducing a finite lifetime, $\tau \sim \Gamma^{-1}$, for the hybridization bands in their full Green's functions. Figure \ref{figs2_1} gives the calculated conductance spectra for $q=0, 0.2, \infty$ and $n_L=1, 16$ with increasing $\Gamma$ from 0.01 to 0.5. We see in all the cases, the finite $\Gamma$ leads to a broadening of the spectra. For $q=0$, the spectra then exhibit a broad minimum and become asymmetric with respect to the bias voltage due to the Van Hove singularity, which could be easily confused with the typical asymmetric Fano lineshape at finite $q$. For $q=0.2$, the sharp "mid-gap" peak is slightly broadened but remains robust for $\Gamma=0.2$. It only disappears at an extremely large $\Gamma=0.5$, where the spectra show a broad peak locating at different energies for the two configurations, reflecting the suppression of the $f$-electron contribution for large $n_L$. The case of $q=\infty$ is more interesting. For $\Gamma=0.2$, the hybridization gap is still seen in the $1\to 1$ configuration, but almost completely smeared out for the $16\times 16\to 16\times 16$ case. The results for large $\Gamma$ at $q=\infty$ resemble that of $q=0$, calling for caution in interpreting the experimental data. We note that the peak for $q=0$ comes from the Van Hove singularity of the 2D lattice, whereas that for $q=\infty$ originates from the $f$-electron spectral weight at the edge of the hybridization gap. Hence interpretation of the experimental data requires very careful analysis and may strongly depend on the strength of electronic correlations. In most of the heavy fermion compounds, experiments suggest an intermediate $\Gamma$ so that the hybridization gap is suppressed in PCS but shows up in STS \cite{CeCoIn5Park2008PRL,CeCoIn5Aynajian}. In URu$_2$Si$_2$, however, it has been argued that "hidden order" may reduce the strength of $\Gamma$, providing a plausible explanation to the observed gap feature in PCS at low temperatures \cite{URu2Si2Schmidt2010,URu2Si2Aynajian2010PNAS,URu2Si2ParkPRL}.

\begin{figure}[t]
  \begin{center}
  \includegraphics[width=\columnwidth]{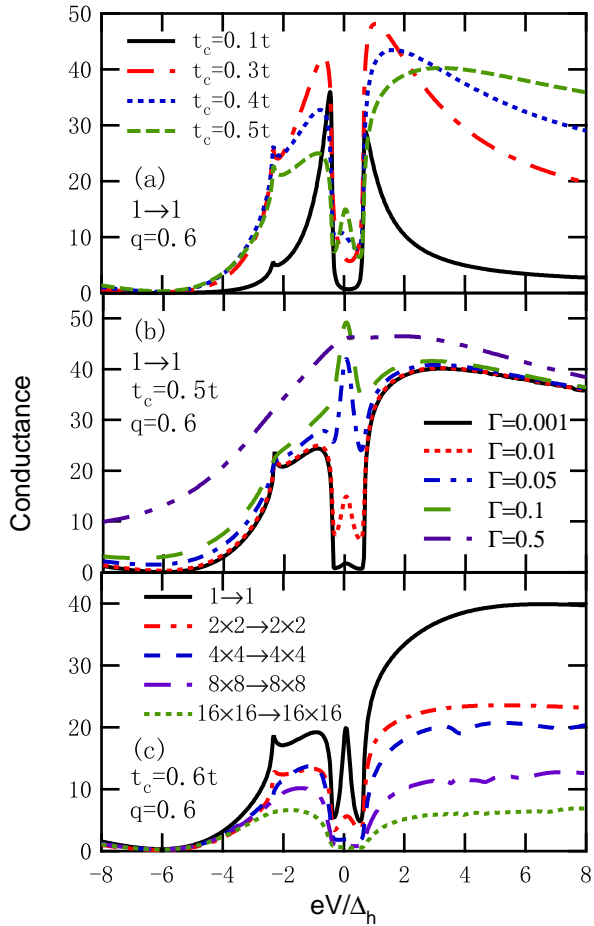}
  \caption{(Color online)  Variation of the conductance spectra with $t_c$, $\Gamma$ and $n_L$ in the strong coupling case. The parameters are $t=25$, $\mu=-20$,  $\epsilon_f=0.01$, $\mathcal{V}=5$, $t_L=t$. In (a,c) $\Gamma=0.01$. The conductance is in units of $\frac{2e^2}{h}n_L^2\frac{4\pi }{t_Lt}$. }
  \label{figs2_2}
  \end{center}
\end{figure}

\subsection{Strong coupling ($t_{f,c}\sim t$)}

Finally, we discuss briefly the strong coupling ($t_{f,c}\sim t$) case. Figure \ref{fig_ns} presents the spectra calculated with increasing $t_c/t$ from 0.1 to 0.5 for $q=0.2$. We see no qualitative change in the spectra, except that the asymmetry of the spectra is somewhat suppressed for larger $t_c$. This effect comes from the second term of the current formula in Eq. (\ref{Eqconductance01}), which contributes a nonequilibrium feedback on the interface and partly suppresses the electron transport. However, further increasing the hopping parameter $t_f$ reveals a small peak inside the hybridization gap, as shown Fig. \ref{figs2_2}(a) for $q=0.6$, or $t_f/V=1.5$, a very strong coupling between the lead and the $f$-electron site. We hence interpret this peak to originate from the impurity effect introduced by the local tip. Mathematically, this also comes from the second term of Eq. (\ref{Eqconductance01}). For the $1\to1$ configuration, the first term is directly related to the density of states or imaginary part of the lattice Green's functions. However, the real part, which diverges at the edge of the hybridization gap, may also have a large counter-contribution in the second term, leading to the suppression of the conductance at the gap edges and hence the "mid-gap" peak. As shown in Fig. \ref{figs2_2}(b), this peak feature is enhanced for moderate $\Gamma$ before it is eventually smeared out. Figure \ref{figs2_2}(c) shows the results for other configurations. We see the peak is gradually suppressed with increasing $n_L$, possibly due to the suppression of the impurity state. We note that the extremely large $t_{c}$ or $t_f$ may not be realistic in real scanning tunneling or point contact measurements, but they may be realized in some special experimental design.

\section{Discussions and Conclusion}
We have presented systematic study of the variation of the conductance spectra as a function of the Fano parameter $q$ and the size of the junction interface $n_L$ for a periodic Anderson lattice. Our results reproduce the well-known Fano interference \cite{YangPRB2009} and reveal new features such as the spatial interference effect for finite interface. We show interesting interplay between both effects and the band features including the hybridization gap and the Van Hove singularity. Our study reflects some of the features of STS at small $n_L$ and PCS at large $n_L$ and shows several essential differences between these two important techniques. First, the PCS differential conductance is modulated by spatial interference effect, which could suppress some of the band features and, in the large $n_L$ limit, impose a constraint on the momentum space. It may therefore miss some $f$-electron spectral weight at the edges of the hybridization gap. As a possible "smoking gun", we would like to propose that in the case of Kondo insulators, PCS or planar junction might overestimate the size of the insulating gap and yield wrong signatures on the insulator-to-metal transition with doping. As a matter of fact, the bulk insulating gap in SmB$_6$ was found to be $\sim18-21\,$meV in PCS and junctions \cite{XHZhangPRX,Flachbart2001PRB,Park2016PNAS} and $\sim16\,$meV in STS \cite{SmB6Hoffman,SmB6PNAS2014}. Our observation is a general result based purely on the momentum and energy conservation in the $n_L=\infty$ limit and may be extended to other materials. For semimetals or topological surface states in which the Fermi pockets locate in a small region of the Brillouin zone, PCS might also miss the states near the Fermi energy and yield a finite insulating gap. For example, for the particular lead in our model, it may miss the Dirac point at $\mathbf{k}=(0,0)$ in the surface Brillouin zone in SmB$_6$. One has to be careful in data analysis. In the case of STS, on the other hand, this is typically not an issue but spatial interference may also occur depending on the location of the tip. In general, PCS measures a weighted average over the Brillouin zone, causing its insensitivity over some electronic features. However, we expect that phase transitions accompanied with abrupt change of the whole Fermi surface should usually be detectable in PCS. Second, in the strong coupling limit, STS may exhibit a mid-gap peak due to impurity effect, while it is not present in PCS. This peak should not be confused with the very exotic Majorana mode at zero bias. Third, PCS often involves many relaxation processes and incoherent scatterings which are not included in this study. These effects will also cause broadening of the PCS spectra and, together with the interference effect, explain the suppression of the hybridization gap feature seen in many experiments. For example, in $\rm{CeCoIn_5}$ hybridization gap is observed in STS \cite{CeCoIn5Aynajian} but not in PCS \cite{CeCoIn5Park2008PRL}.

The realistic situations are of course more complicated. In point contact experiments, the junction between the lead and the lattice may contain hundreds of small contacts at the interface with different hopping configurations and different coupling strengths, so the PCS may only be understood as a statistical average of these small contacts.
We have shown that the spatial interference effect strongly depends on the particular configuration of the interface. However, the results are qualitatively unchanged for large $n_L$. We hence expect that our toy model captures some truth of the fundamental physics involving the electron transport, such as the Fano interference due to the existence of both the conduction and $f$-electron channels and the spatial interference effect due to the finite size of the junction. We further note that the mean-field treatment of the periodic Anderson lattice may also miss a lot of interesting physics in a heavy fermion system, such as the effect of the quantum criticality. A complete treatment also needs to take into account some other details including self-consistent modifications on the local hybridization gap. This is, however, beyond the purpose of our toy model to give a simple illustration of the underlying physical mechanism of the electron transport alone. We therefore leave it for future investigations. Nevertheless, our observation of the interesting interplay between these interference effects and the band features such as the Van Hove singularity and the hybridization gap provides new insights into the electron transport of heavy-fermion materials. Most importantly, we show based on general principles that PCS and planar junction may not be able to cover all the electronic states in the Brillouin zone. These call for caution in interpreting the experimental data in realistic measurement.

\section*{ACKNOWLEDGEMENTS}
This work was supported by the State Key Development Program for Basic Research of China (Grant No. 2015CB921303), the National Natural Science Foundation of China (NSFC Grants No. 11522435, No. 11176002), the National High Technology Research and Development Program of China (Grant No. 2015AA01A304) and the Strategic Priority Research Program of the Chinese Academy of Sciences (Grant No. XDB07020200).

\begin{appendix}

\section{}
\label{appendixCformula}
Here we derive Eq. (\ref{EqGle}) for the lesser Green's function of the lattice electrons coupled with the lead. Using the Dyson equation shown in Eq. (\ref{EqGg}), we have
\begin{align}
&\check G_{SS}^{r}-\check G_{SS}^{a}=-\check G_{SS}^{r}(\check G_{SS}^{r,-1}-\check G_{SS}^{a,-1})\check G_{SS}^{a}\nonumber\\
&=-\check G_{SS}^{r}(\check g_{SS}^{r-1}-\check g_{SS}^{a-1})\check G_{SS}^{a}+\check G_{SS}^{r}(\check \Sigma_{L}^{r}-\check \Sigma_{L}^{a})\check G_{SS}^{a}.
\end{align}
Using the fluctuation-dissipation theorem for the lattice electrons, $\check g_{SS}^<=-f_S(\check g_{SS}^{r}-\check g_{SS}^{a})$, we can multiply the above formula with $-f_S$ on both sides to give
\begin{eqnarray}
-f_{S}(\check G_{SS}^{r}-\check G_{SS}^{a})=&\check G_{SS}^{r}\check g_{SS}^{r-1}\check g_{SS}^{<}\check g_{SS}^{a-1}\check G_{SS}^{a}\nonumber\\
&-f_{S}\check G_{SS}^{r}(\check \Sigma_{L}^{r}-\check \Sigma_{L}^{a})\check G_{SS}^{a}.
\end{eqnarray}
On the other hand, we have, using the Keldysh equation,
\begin{eqnarray}
\check G_{SS}^<=\check G_{SS}^r\check g_{SS}^{r-1}\check g_{SS}^<\check g_{SS}^{a-1}\check G_{SS}^a+\check G_{SS}^r\check \Sigma_{L}^{<}\check G_{SS}^a.
\end{eqnarray}
Combining the above equations yield,
\begin{eqnarray}
\check G_{SS}^{<}=-f_{S}(\check G_{SS}^{r}-\check G_{SS}^{a})+\check G_{SS}^{r}\check \Sigma_{L}^{<}\check G_{SS}^{a}\nonumber\\
+f_{S}\check G_{SS}^{r}(\check \Sigma_{L}^{r}-\check \Sigma_{L}^{a})\check G_{SS}^{a}.
\end{eqnarray}
Using once again the fluctuation-dissipation theorem for the self-energy of the electrons in the lead, $\check \Sigma_L^<=if_L(\check \Sigma_L^r-\check \Sigma_L^a )=if_L\check \Gamma_L$, we derive Eq. (\ref{EqGle}) in the main text for the nonequilibrium lesser Green's function of the lattice electrons,
\begin{eqnarray}
\check G_{SS}^<=-\!f_S(\check G_{SS}^r-\!\check G_{SS}^a)
-i(f_S-f_L)\check G_{SS}^r\check \Gamma_{L}\check G_{SS}^a.
\end{eqnarray}

\section{}
\label{appendixScattering}
Here we prove that our current formula in Eq. (\ref{currentformula1}) with the additional higher-order term is consistent with the well-known Landauer formula \cite{book(Datta)}. To show this, we divide the whole system into three parts: the left lead ($L$), the central part containing the lattice sites at the junction, and the rest of the lattice as the right lead ($S$). The two leads are considered to be in equilibrium state with the Fermi distribution function of $f_L$ and $f_S$, while the central part is in the nonequilibrium state.

We first rewrite our current formula following the general method in the scattering theory using the scattering matrices, $\check S_{\alpha \beta}$, where $\alpha,\beta=L,S$ denote all the transport channels. The scattering matrices and the Green's functions of the central part ($\check G^{r}$) are related by the so-called Fisher-Lee relation \cite{FisherLee,FisherLeeButtiker},
\begin{eqnarray}
\check S_{\alpha \beta } = \check \delta _{\alpha \beta }-i2\pi\check W_{\alpha }^{\dag
}\check G^{r}\check W_{\beta },
\label{EqFisherLee}
\end{eqnarray}
in which $\check W_\alpha$ satisfies $\check \Gamma _{\alpha} =2\pi W_{\alpha }W_{\alpha }^{\dag }$ and $\check \delta_{\alpha\beta}$ is a unit matrix so that ${\rm Tr}\check \delta_{\alpha\alpha}$ is given by their respective number of transport channels. We have ${\rm Tr}\check \delta_{LL}=n_L^2$ and ${\rm Tr}\check \delta_{LS}=0$. The two terms in Eq. (\ref{currentformula1}) can be rewritten as,
\begin{eqnarray}
\!\!&&{\rm Tr}[i\check \Gamma _{L}(\check G^{r}-\check G^{a})]
={\rm Tr}[(\check \delta _{LL}-\check S_{LL})+(\check \delta_{LL}-\check S_{LL}^{\dag})],\nonumber\\
\!&&{\rm Tr}[\check \Gamma_{L}\check G^{r}\check \Gamma _{L}\check G^{a}]
\!\!=\!{\rm Tr}[(\check \delta_{LL}-\check S_{LL})(\check \delta_{LL}-\check S_{LL}^{\dag})].
\label{EqFisherLee01}
\end{eqnarray}
On the other hand, the current conversation requires the scattering matrices to be unitary,
\begin{equation}
{\rm Tr }\check \delta _{LL}={\rm Tr }(\check S_{LL}\check S_{LL}^{\dag }+\check S_{LS}\check S_{SL}^{\dag }).
\end{equation}
Combining the above equations immediately gives,
\begin{eqnarray}
I=\frac{2e}{h}\int d\omega (f_{L}-f_{S}) T_{LS},
\label{Landauerformula}
\end{eqnarray}
in which $T_{LS}={\rm Tr }[\check S_{LS}\check S_{SL}^{\dag }]$ is the transmission coefficient. This is the famous Landauer formula \cite{book(Datta)}. We see that the second term in Eq. (\ref{currentformula1}) is necessary for the consistency.

\section{}
\label{appendixGreen}
In this appendix we give the explicit form of lattice Green's functions in momentum space which are used in the discussion of the Fano interference \cite{book(ZZLi)}:
\begin{eqnarray}
\check g_{cc}^{r}({\mathbf k},\omega)&=&\frac{1}{(\omega-\frac{ {\mathcal V}^{2}}{\omega-\epsilon_{f}})-\epsilon_{c\mathbf{k}}+i0^{+} },\nonumber\\
\check g_{ff}^{r}({\mathbf k},\omega)&=&\frac{{\mathcal V}^{2}}{(\omega-\epsilon_{f}+i0^{+})^{2}}\check g_{cc}^{r}({\mathbf k},\omega) +\frac{1}{\omega-\epsilon_{f}+i0^{+}},\nonumber\\
\check g_{cf}^{r}({\mathbf k},\omega)&=&\frac{{\mathcal V}}{\omega-\epsilon_{f}}\check g_{cc}^{r}({\mathbf k},\omega).
\end{eqnarray}
Since the prefactor is independent of the momentum, it is clear that the last two equations also hold in real space. We have then,
\begin{eqnarray}
\check\rho_{cf}(\omega)&=&\frac{{\mathcal V}}{\omega-\epsilon_f}\check\rho_c(\omega),\nonumber\\
\check\rho_f(\omega)&=&\frac{{\mathcal V}^2}{(\omega-\epsilon_f)^2}\check\rho_c(\omega),\\ \nonumber
\label{Eqrho}
\end{eqnarray}
which, combined with Eq. (\ref{Eqconductance03}), immediately yields Eq. (\ref{Eqconductance04}) in the main text.

\end{appendix}

\end{document}